\documentclass[12pt]{article}
\begin{document}
      \sloppy

\def\AFOUR{%
\setlength{\textheight}{9.0in}%
\setlength{\textwidth}{5.75in}%
\setlength{\topmargin}{-0.375in}%
\hoffset=-.5in%
\renewcommand{\baselinestretch}{1.17}%
\setlength{\parskip}{6pt plus 2pt}%
}
\AFOUR
\def\car{\mathop{\square}}
\def\carre#1#2{\raise 2pt\hbox{$\scriptstyle #1$}\car_{#2}}

\parindent=0pt
\makeatletter
\def\section{\@startsection {section}{1}{\z@}{-3.5ex plus -1ex minus
   -.2ex}{2.3ex plus .2ex}{\large\bf}}
\def\subsection{\@startsection{subsection}{2}{\z@}{-3.25ex plus -1ex minus
   -.2ex}{1.5ex plus .2ex}{\normalsize\bf}}
\makeatother
\makeatletter
\@addtoreset{equation}{section}
\renewcommand{\theequation}{\thesection.\arabic{equation}}
\makeatother

\renewcommand{\a}{\alpha}
\renewcommand{\b}{\beta}
\newcommand{\g}{\gamma}           \newcommand{\G}{\Gamma}
\renewcommand{\d}{\delta}         \newcommand{\D}{\Delta}
\newcommand{\e}{\varepsilon}
\newcommand{\la}{\lambda}        \newcommand{\LA}{\Lambda}
\newcommand{\m}{\mu}
\newcommand{\A}{\widehat{A}^{\star a}_{\mu}}
\newcommand{\Ar}{\widehat{A}^{\star a}_{\rho}}
\newcommand{\n}{\nu}
\newcommand{\om}{\omega}         \newcommand{\OM}{\Omega}
\newcommand{\p}{\psi}             \newcommand{\PS}{\Psi}
\renewcommand{\r}{\rho}
\newcommand{\s}{\sigma}           \renewcommand{\S}{\Sigma}
\newcommand{\f}{{\phi}}           \newcommand{\F}{{\Phi}}
\newcommand{\vf}{{\varphi}}
\newcommand{\y}{{\upsilon}}       \newcommand{\Y}{{\Upsilon}}
\newcommand{\z}{\zeta}

\renewcommand{\AA}{{\cal A}}
\newcommand{\BB}{{\cal B}}
\newcommand{\CC}{{\cal C}}
\newcommand{\DD}{{\cal D}}
\newcommand{\EE}{{\cal E}}
\newcommand{\FF}{{\cal F}}
\newcommand{\GG}{{\cal G}}
\newcommand{\HH}{{\cal H}}
\newcommand{\II}{{\cal I}}
\newcommand{\JJ}{{\cal J}}
\newcommand{\KK}{{\cal K}}
\newcommand{\LL}{{\cal L}}
\newcommand{\MM}{{\cal M}}
\newcommand{\NN}{{\cal N}}
\newcommand{\OO}{{\cal O}}
\newcommand{\PP}{{\cal P}}
\newcommand{\QQ}{{\cal Q}}
\renewcommand{\SS}{{\cal S}}
\newcommand{\RR}{{\cal R}}
\newcommand{\TT}{{\cal T}}
\newcommand{\UU}{{\cal U}}
\newcommand{\VV}{{\cal V}}
\newcommand{\WW}{{\cal W}}
\newcommand{\XX}{{\cal X}}
\newcommand{\YY}{{\cal Y}}
\newcommand{\ZZ}{{\cal Z}}

\newcommand{\ch}{\widehat{C}}
\newcommand{\gh}{\widehat{\gamma}}
\newcommand{\W}{W_{i}}
\newcommand{\na}{\nabla}
\newcommand{\xint}{\dint d^4x\;}
\newcommand{\sla}{\raise.15ex\hbox{$/$}\kern -.57em}
\newcommand{\Sla}{\raise.15ex\hbox{$/$}\kern -.70em}
\def\h{\hbar}
\def\Lp{\displaystyle{\biggl(}}
\def\Rp{\displaystyle{\biggr)}}
\def\LP{\displaystyle{\Biggl(}}
\def\RP{\displaystyle{\Biggr)}}
\newcommand{\lp}{\left(}\newcommand{\rp}{\right)}
\newcommand{\lc}{\left[}\newcommand{\rc}{\right]}
\newcommand{\lac}{\left\{}\newcommand{\rac}{\right\}}
\newcommand{\identity}{\bf 1\hspace{-0.4em}1}
\newcommand{\complex}{{\kern .1em {\raise .47ex
\hbox {$\scriptscriptstyle |$}}
      \kern -.4em {\rm C}}}
\newcommand{\real}{{{\rm I} \kern -.19em {\rm R}}}
\newcommand{\rational}{{\kern .1em {\raise .47ex
\hbox{$\scripscriptstyle |$}}
      \kern -.35em {\rm Q}}}
\renewcommand{\natural}{{\vrule height 1.6ex width
.05em depth 0ex \kern -.35em {\rm N}}}
\newcommand{\tint}{\int d^4 \! x \, }
\newcommand{\intg}{\int d^D \! x \, }
\newcommand{\intm}{\int_\MM}
\newcommand{\tr}{{\rm {Tr} \,}}
\newcommand{\half}{\dfrac{1}{2}}
\newcommand{\pa}{\partial}
\newcommand{\pad}[2]{{\frac{\partial #1}{\partial #2}}}
\newcommand{\fud}[2]{{\frac{\delta #1}{\delta #2}}}
\newcommand{\dpad}[2]{{\displaystyle{\frac{\partial #1}{\partial
#2}}}}
\newcommand{\dfud}[2]{{\displaystyle{\frac{\delta #1}{\delta #2}}}}
\newcommand{\dfrac}[2]{{\displaystyle{\frac{#1}{#2}}}}
\newcommand{\dsum}[2]{\displaystyle{\sum_{#1}^{#2}}}
\newcommand{\dint}{\displaystyle{\int}}
\newcommand{\eg}{{\em e.g.,\ }}
\newcommand{\Eg}{{\em E.g.,\ }}
\newcommand{\ie}{{{\em i.e.},\ }}
\newcommand{\Ie}{{\em I.e.,\ }}
\newcommand{\nb}{\noindent{\bf N.B.}\ }
\newcommand{\etal}{{\em et al.}}
\newcommand{\etc}{{\em etc.\ }}
\newcommand{\via}{{\em via\ }}
\newcommand{\cf}{{\em cf.\ }}
\newcommand{\twiddle}{\lower.9ex\rlap{$\kern -.1em\scriptstyle\sim$}}
\newcommand{\qed}{\vrule height 1.2ex width 0.5em}
\newcommand{\grad}{\nabla}
\newcommand{\bra}[1]{\left\langle {#1}\right|}
\newcommand{\ket}[1]{\left| {#1}\right\rangle}
\newcommand{\vev}[1]{\left\langle {#1}\right\rangle}

\newcommand{\equ}[1]{(\ref{#1})}
\newcommand{\eq}{\begin{equation}}
\newcommand{\eqn}[1]{\label{#1}\end{equation}}
\newcommand{\eea}{\end{eqnarray}}
\newcommand{\eqa}{\begin{eqnarray}}
\newcommand{\eqan}[1]{\label{#1}\end{eqnarray}}
\newcommand{\ba}{\begin{array}}
\newcommand{\ea}{\end{array}}
\newcommand{\eqac}{\begin{equation}\begin{array}{rcl}}
\newcommand{\eqacn}[1]{\end{array}\label{#1}\end{equation}}
\newcommand{\qq}{&\qquad &}
\renewcommand{\=}{&=&} 
\newcommand{\cb}{{\bar c}}
\newcommand{\mn}{{\m\n}}
\newcommand{\pic}{$\spadesuit\spadesuit$}
\newcommand{\?}{{\bf ???}}
\newcommand{\Tr }{\mbox{Tr}\ }
\newcommand{\adot}{{\dot\alpha}}
\newcommand{\bdot}{{\dot\beta}}
\newcommand{\gdot}{{\dot\gamma}}

\global\parskip=4pt
\titlepage  \noindent
{
   \noindent

\hfill GEF-TH-10/2005 


\vspace{2cm}

\noindent
{\bf
{\large  Noncommutative Two Dimensional BF Model 
}}

\vspace{.5cm}
\hrule

\vspace{2cm}

\noindent
{\bf 
Alberto Blasi, Nicola Maggiore and Michele 
Montobbio}

\noindent
{\footnotesize {\it
 Dipartimento di Fisica -- Universit\`a di Genova --
via Dodecaneso 33 -- I-16146 Genova -- Italy and INFN, Sezione di
Genova 
} }

\vspace{2cm}
\noindent
{\tt Abstract~:}
We consider the noncommutative extension of the BF theory in two 
spacetime dimensions. We show that the introduction of the 
noncommutative parameter $\theta_{\m\n}$, already at first order 
in the analytical 
sector, induces infinitely many terms in the quantum extension of the 
model. This clashes with the commonly accepted rules of QFT, and we 
believe that this 
problem is not peculiar to this particular model, but it might concern 
the noncommutative extension of any ordinary quantum field theory 
obtained via the Moyal prescription. A detailed study of noncommutative 
anomalies is also presented.
\vfill\noindent
{\footnotesize {\tt Keywords:}
Noncommutative field theory, Algebraic renormalization, BRS
quantization\\
{\tt PACS Nos:} 11.10.Gh Renormalization, 
          11.10.Nx Noncommutative field theory,
          11.15.-q Gauge field theories.}
\newpage
\begin{small}
\end{small}

\setcounter{footnote}{0}


\section{Introduction}

The literature on noncommutative field theory models has rapidly 
grown to a considerable size, and yet there is no clear cut 
recipe of how one should proceed to analyze such models in particular 
for what concerns their renormalizability. One of the main issues 
which arise, is whether the noncommutative model, with 
$\theta_{\m\n}$ as the noncommutative parameter, has a smooth 
(analytical) $\theta_{\m\n}\rightarrow 0$ limit or not. In the first case 
we have a corresponding standard, commutative model, which can be 
treated with the well established machinery of local quantum field 
theory, and it is tempting to apply it also to the noncommutative 
model 
\cite{Becchi:2003dg,Barnich:2003wq,Vassilevich:2004ym,Soroush:2003pk}. 
In the second case, the model we are looking at is 
intrinsically noncommutative, and we loose both analyticity in 
$\theta_{\m\n}$ and locality 
\cite{Wulkenhaar:2001sq,Blaschke:2005dv}. 

However, we would like to emphasize 
that in any approach the theory should be defined by the functional 
equations which encode its symmetry content and all the additional 
information needed to analyze both the stability and the anomaly 
issues. Of course those tools are normally applied in an environment 
where locality, power counting and the Quantum Action Principle 
\cite{Lowenstein:1971vf,Lowenstein:1971jk,Lam:1972mb,Lam:1973qa,Clark:1976ym}
also 
hold true; we might call it a traditional reference frame. 
In the 
study of the noncommutative extensions of a standard quantum field 
theory model, is a matter of personal choice how much we should adhere 
to this traditional reference frame. 

In this paper we would like to 
probe where we are led to by analyzing the analytical $\theta_{\m\n}$ 
sector of the two dimensional BF model 
\cite{Birmingham:1991ty,Blau:1989bq,Blasi:1992hq}, using the entire machinery of 
the traditional reference frame \cite{Piguet:1995er}. 
Our idea is that
the study of
noncommutative models beginning from a commutative one, looking only
at the sector analytical in the $\theta^{\mu\nu}$ tensor, is of interest 
since it can provide hints to new and unexpected features of the
noncommutative theory.  Here we provide an example of problems which
arise in carrying out the process of an analytical $\theta^{\mu\nu}$
expansion. In other words, we limit ourselves to the sector of the
theory in which the UV properties are controlled by the planar
diagrams \cite{Becchi:2003dg,Douglas:2001ba}. Our aim is to show 
that this sector,
in which the Quantum Action Principle should reign,
does not admit a consistent quantum extension.

When attempting to define a noncommutative quantum field theory
\cite{Szabo:2001kg} and wishing also to arrive at a formulation which  
allows explicit amplitude computation, one is faced with the problem of 
choosing a precise form for the non 
commutative product. One of the most popular choices is the 
Groenewold-Moyal product
\cite{Groenewold:1946kp,Moyal:1949sk}
which is implemented with a simple 
exponential formula and needs the introduction of an antisymmetric 
constant tensor $\theta^{\mu\nu}$ having the 
dimensions of an inverse mass squared. It is commonly accepted 
that this procedure leads to a well defined 
noncommutative theory if the commutative model we begin with is 
sound. We shall show that this is not always the case
by providing a counterexample. 

A reliable noncommutative extension of an ordinary model should be based on the functional 
identities encoding the symmetries, 
on locality and 
power counting, just as it happens in the commutative case. 
This procedure, in the standard case, leads to the stability
and anomaly analysis i.e. the model is perturbatively renormalizable 
if the classical action is the most general local 
functional compatible with the above constraints (stability) and the 
symmetries are not broken by the radiative 
corrections (anomaly) \cite{Piguet:1995er}. 

The paper is organized as follows. To fix the notations, 
in Section 2 we briefly recall
the functional equations 
(BRS identity, Landau gauge, ghost equation and vector
supersymmetry) which form a closed algebraic structure and completely 
define, together with locality and power counting,
the commutative model \cite{Blasi:1992hq}. 
In Section 3 we deform the classical theory by means of the 
anticommuting constant parameter $\theta_{\m\n}$, introduced by means 
of the substitution of the ordinary product with the 
Groenewold-Moyal star product. 
In \cite {Blasi:2005bk} we have showed that
the defining equations remain exactly the same we have in 
the commutative case. Of course this is not so for the
classical action which acquires, at the first order in 
$\theta^{\mu\nu}$, a local contribution with canonical
dimension equal to four and coupled to $\theta^{\mu\nu}$ itself.
In Section 4 we are led to consider the stability problem of the classical 
action to first order in  $\theta^{\mu\nu}$, {\it i.e.} we completely 
characterize the counterterm, and conclude that it contains a 
countable infinity of terms linked to two classes of 
coupling constants.  Finally, in Section 5, we also analyzed 
the anomaly problem and 
found that, at least to first order in  $\theta^{\mu\nu}$, the model 
is anomaly free. 
We draw our conclusions in the final Section 6.

\section{The commutative model}

The action of ordinary, commutative, $BF$ model in two spacetime
dimensions is \cite{Birmingham:1991ty,Blau:1989bq,Blasi:1992hq}
\eq
S_{inv} =
-\half\Tr \int d^{2}x\ \epsilon^{\mu\nu} F_{\mu\nu} \phi 
=
-\half\int d^{2}x\ \epsilon^{\mu\nu}  F^{a}_{\mu\nu} \phi^{a}
\eqn{2.1}
where $\epsilon^{\mu\nu}$ is the two-dimensional completely
antisymmetric Levi-Civita tensor, 
$F_{\mu\nu}= F^{a}_{\mu\nu} T^{a} =
\partial_{\mu}A_{\nu} - \partial_{\nu}A_{\mu}
-i[A_{\mu},A_{\nu}]$ 
is the field strength, and $A_{\mu}(x) = A^{a}_{\mu}(x)T^{a}$ and
$\phi(x)=\phi^{a}(x)T^{a}$ are the gauge field and a scalar field
respectively, belonging to the adjoint representation of
the gauge group, assumed to be $U(n)$, whose generators $T^{a}\
(a=1,\ldots, n^{2})$ obey
\begin{eqnarray}
    \Tr(T^{a}T^{b}) &=& \delta^{ab} \label{2.2}\\
    T^{a}T^{b} &=& \frac{i}{2} f^{abc}T^{c} +\frac{1}{2} d^{abc}
    T^{c}\label{2.3}
    \end{eqnarray}
$f^{abc}$ and $d^{abc}$ being the antisymmetric and symmetric Gell-Mann
tensors, respectively. We will comment on this choice of the gauge group 
later. 

The action \equ{2.1} is  invariant under the following
infinitesimal gauge transformations, with $\Lambda(x)=\Lambda^{a}(x)T^{a}$
as infinitesimal gauge parameter
\begin{eqnarray}
    \delta_{g} A_{\mu} &=& D_{\mu} \Lambda = \partial_{\mu} \Lambda   
    -i[A_{\mu},\Lambda] \label{2.4}\\
    \delta_{g}\phi &=& -i[\phi,\Lambda]\label{2.5}
    \end{eqnarray}
As usual, the quantization of the model proceeds with the introduction 
of a set of quantum fields $(c^{a}(x),\bar{c}^{a}(x),b^{a}(x))$,
playing the role of ghost, antighost and Lagrange multiplier fields,
respectively, by means of which the gauge fixing term reads, in the
Landau gauge,
\eq
S_{gf} = \int d^{2}x\ \left (b^{a}\partial^{\mu}A^{a}_{\mu} -
\bar{c}^{a}\partial^{\mu}(D_{\mu}c)^{a}\right )
\eqn{2.6}

The transformations \equ{2.4}-\equ{2.5} are 
no more a symmetry of the gauge fixed
action $S = S_{inv}[A] + S_{gf}[A,c,\bar{c},b]$, which is indeed 
invariant under the
BRS transformations
\begin{eqnarray}
    s A_{\mu} &=& D_{\mu}c \nonumber \\
    s \phi &=& -i[\phi,c] \nonumber\\
    s c &=& i c^{2} \label{2.7}\\
    s \bar{c} &=& b \nonumber\\
    s b &=&  0\nonumber
    \end{eqnarray}
The BRS operator $s$ is characterized by the two fundamental
properties of being nilpotent
\eq
s^{2} =0
\eqn{2.8}
and of being a symmetry of the theory
\eq
s S = 0
\eqn{2.9}
The latter point is most easily verified once it is noticed that the
gauge fixing term $S_{gf}$ can be written as a BRS cocycle
\eq
S_{gf} = s \int d^{2}x\ \bar{c}^{a}\partial^{\mu}A^{a}_{\mu}
\eqn{2.10}
Since the BRS transformations \equ{2.7} are nonlinear, external fields
${A^{*}}^{a\mu}(x)$, ${\phi^{*}}^{a}(x)$ and ${c^{*}}^{a}(x)$ must 
be coupled to
the nonlinear BRS variations through
\eq
S_{ext}=\int d^{2}x\  \left ( {A^{*}}^{a\mu}(s A^{a}_{\mu}) +
{\phi^{*}}^{a}(s\phi^{a}) + {c^{*}}^{a}(sc^{a})\right )
\eqn{2.11}
in order to be able to write for the total classical action
\eq
\S=S_{inv} +S_{gf}+S_{ext}
\eqn{2.12}
a Slavnov-Taylor identity
\eq
{\cal S}(\Sigma) = \Tr \int d^{2}x \left (
\fud{\Sigma}{A^{*\mu}} \fud{\S}{A_{\mu}} +
\fud{\Sigma}{\phi^{*}} \fud{\S}{\phi} +
\fud{\Sigma}{c^{*}} \fud{\S}{c}
+ b\fud{\S}{\bar{c}}
\right ) =0\ .
\eqn{2.13}
The action $S$ is topological, since it does not depend on the
spacetime metric $g_{\mu\nu}$. In other words, only the gauge fixing
term of the action contributes to the energy-momentum tensor, which
therefore is an
exact BRS cocycle
\eq
T_{\mu\nu}\equiv\fud{S}{g^{\mu\nu}} = s \Lambda_{\mu\nu}
\eqn{2.14}
for some integrated local functional $\Lambda_{\mu\nu}$, which can be easily
calculated from the expression of the classical action $S$. Now, as it
has been remarked for the first time in \cite{Maggiore:1991aa}, 
the highly non trivial
observation that both $T_{\mu\nu}$ and $\Lambda_{\mu\nu}$ are
conserved, underlies the existence of an additional  symmetry
of the action. Indeed, the conservation relation
\eq
\partial^{\nu}\Lambda_{\mu\nu} = \mbox{contact terms}
\eqn{2.15}
once integrated, directly represents the Ward identity for the vector symmetry
\begin{eqnarray}
\delta_{\mu}A_{\nu} &=& 0 \nonumber\\
\delta_{\mu}\phi &=&   \epsilon_{\mu\nu}\partial^{\nu}\bar{c}  \nonumber\\
\delta_{\mu}c &=&  A_{\mu}\label{2.16}\\
\delta_{\mu}\bar{c}  &=& 0\nonumber\\
\delta_{\mu}b &=&   \partial_{\mu}\bar{c} \nonumber
\end{eqnarray}
The existence of such a linear, vector symmetry is peculiar to topological
field theories, and it is called vector $supersymmetry$ due to the
following algebra, formed by the operators $s$ and $\delta_{\mu}$
\begin{eqnarray}
    s^{2}&=&0 \nonumber\\
    \{\delta_{\mu},\delta_{\nu}\} &=& 0 \label{2.17}\\
    \{\delta_{\mu},s\} &=& \partial_{\mu} + \mbox{eqs of 
    motion}\nonumber
    \end{eqnarray}
which, closing on shell on the spacetime translations, describe a 
superalgebra of the Wess-Zumino type. Once the source term 
\equ{2.11} 
is introduced, the algebra \equ{2.17} closes off shell, and the Ward
identity for the vector supersymmetry becomes
\eq
W_{\mu}\S = \Delta_{\mu}
\eqn{2.18}
where
\eq
W_{\mu}=\Tr \int d^{2}x\
\left(
\epsilon_{\mu\nu}(\partial^{\nu}\bar{c} + A^{*\nu})\fud{}{\phi} +
\epsilon_{\mu\nu}\phi^{*}\fud{}{A_{\n}} +
A_{\mu}\fud{}{c} +
\partial_{\mu}\bar{c}\fud{}{b} -
c^{*}\fud{}{A^{*\mu}}
\right ) 
\eqn{2.19}
and  $\Delta_{\mu}$ is a breaking linear in the quantum fields, and
therefore purely classic
\eq
\Delta_{\mu} = \Tr \int d^{2}x\
\left ( -A^{*\nu}\partial_{\mu}A_{\nu} 
- \phi^{*}\partial_{\mu}\phi
+c^{*}\partial_{\mu}c
+\epsilon_{\mu\nu}\phi^{*}\partial^{\nu}\bar{c}
\right)
\eqn{2.20}

In addition, as for any other - commutative - gauge field, the 
choice of the Landau gauge guarantees the existence of an additional
constraint on the classical action $\Sigma$: the ``ghost'' equation
\cite{Blasi:1990xz}
\eq
{\cal G}^{a}\Sigma = \int d^{2}x \
\left (
\fud{}{c^{a}} + f^{abc} \bar{c}^{b}\fud{}{b^{c}}
\right ) \Sigma = \Delta^{a}\ ,
\eqn{2.21}
where $\Delta^{a}$ is again a classical breaking
\eq
\Delta^{a} = \int d^{2}x\
f^{abc} \left (
A^{*b\mu}A^{c}_{\mu} + \phi^{*b}\phi^{c} - c^{*b}c^{c} \right )\ .
\eqn{2.22}

The following constraints 
\begin{enumerate}
    \item the Slavnov-Taylor identity \equ{2.13};
    \item the supersymmetry \equ{2.18};
    \item the ghost equation \equ{2.21}
    \item the Landau gauge condition
    \eq
    \fud{\Sigma}{b^{a}} = \partial^{\mu}A^{a}_{\mu}
    \eqn{2.23}
    \end{enumerate}
fully characterize the theory, both at the classical and at the
quantum level. As showed in \cite{Blasi:1992hq}, the classical action $\Sigma$ 
\equ{2.12}
is the most general one compatible with the whole set of constraints, 
which in turn are free of anomalies.

In other words, the ordinary, commutative BF theory in two spacetime
dimensions is renormalizable to all orders of perturbation theory. 
More than this, we know that this theory, like any other topological 
quantum field theory, is finite, namely no radiative corrections are 
allowed by the symmetries characterizing the classical theory.

\section{The noncommutative model}

The usual and generally accepted behavior 
to proceed towards the noncommutative extension of
any commutative field theory, is simply to substitute the ordinary
product between fields with the Moyal ``star'' product 
\cite{Groenewold:1946kp,Moyal:1949sk}
\eq
\phi(x)\psi(x) \longrightarrow \phi(x)*\psi(x) \equiv
\lim_{y\rightarrow x}\
\exp(\frac{i}{2}\theta^{\mu\nu}\partial^{x}_{\mu}\partial^{y}_{\nu})\
\phi(x)\psi(y)\ ,
\eqn{3.1}
where $\theta^{\mu\nu}$ is a rank-two antisymmetric matrix which 
controls the noncommutative nature of spacetime coordinates
\eq
[x^{\mu},x^{\nu}]=i\theta^{\mu\nu}
\eqn{3.2}
The choice of $U(n)$ as gauge group is 
motivated by the request of having gauge-valued noncommutative 
fields, which would not be the case for a generic else non abelian gauge 
group, for \equ{2.3} acquires an additional term
\eq
 T^{a}T^{b} = \frac{1}{2n} \delta^{ab} 
 + \frac{i}{2} f^{abc}T^{c} +\frac{1}{2} d^{abc}
    T^{c}
\eqn{3.3}
As a consequence of the absence of the 
central term $\frac{1}{2n} \delta^{ab}$, the 
gauge group $U(n)$ is closed under the star product, while, for 
instance, $SU(n)$ does not give rise to any gauge group on 
the noncommutative plane.

In two spacetime dimensions, the commutator \equ{3.2} is Lorentz
invariant. This implies that $\theta^{\mu\nu}$ is proportional to the 
Levi-Civita tensor $\epsilon^{\mu\nu}$
\eq
\theta^{\mu\nu} = \frac{1}{\theta^{2}}\ \epsilon^{\mu\nu}
\eqn{3.4}
where $\theta^{2}$ is the dimensional parameter induced by the commutation 
relation \equ{3.2}. The property \equ{3.4} is tightly related to the two
dimensional spacetime.

We will not bother the reader with the motivations for such a
generalization of ordinary spacetime, and of its consequences on field
theory, for which an excellent literature exists 
\cite{Szabo:2001kg,Douglas:2001ba}.

In this paper, we simply investigate the consequences of the
introduction in the theory of $\theta^{\mu\nu}$ as a new
ingredient in an otherwise ordinary gauge field theory. Under this
respect, $\theta^{\mu\nu}$ is no more than an antisymmetric
constant parameter with negative (minus two) mass dimensions, coupling
to fields to form monomials. In particular,  in the example of
two dimensional BF theory we ask if the deformation induced by the 
presence of the noncommutative parameter $\theta^{\m\n}$
is compatible with the basic features of quantum field
theory, namely
\begin{itemize}
\item stability of the classical action
\item absence of anomalies
\end{itemize}
The field strength becomes
\begin{eqnarray}
F_{\mu\nu}(\theta) &=& \partial_{\mu}A_{\nu} - \partial_{\nu}A_{\mu}
- ig (A_{\mu}*A_{\nu} - A_{\nu}*A_{\mu}) \nonumber \\
&=& \partial_{\mu}A_{\nu} - \partial_{\nu}A_{\mu}
-ig [A_{\mu},A_{\nu}]_{*}\label{3.5}
\end{eqnarray}
and, at the first order in $\theta$
\eq
F^{a}_{\mu\nu}(\theta) = F^{a}_{\mu\nu}(0)
+\frac{1}{2}d^{abc}\theta^{\alpha\beta}
(\partial_{\alpha}A^{b}_{\mu})(\partial_{\beta}A^{c}_{\nu})
\eqn{3.6}
Analogously, the covariant derivative, at the first order in $\theta$,
reads
\eq
(D_{\mu}\Lambda)^{a}(\theta) = (D_{\mu}\Lambda)^{a}(0)
+\frac{1}{2}d^{abc}\theta^{\alpha\beta}
(\partial_{\alpha}A^{b}_{\mu})(\partial_{\beta}\Lambda^{c})
\eqn{3.7}
The invariant, gauge-fixed, $\theta$-dependent action is
\begin{eqnarray}
S^{(\theta)} &=& \Tr \int d^{2}x\
\left( -\frac{1}{2}\epsilon^{\mu\nu}F_{\mu\nu}(\theta)*\phi + s^{(\theta)}
(\bar{c}\partial A)\right)\label{3.8}  \\
&=& S + \frac{1}{2}d^{abc}\theta^{\alpha\beta} \int d^{2}x\
\left (
-\frac{1}{2}\epsilon^{\mu\nu}
\phi^{a}(\partial_{\alpha}A^{b}_{\mu})(\partial_{\beta}A^{c}_{\nu})
+
(\partial_{\mu}\bar{c}^{a})(\partial_{\alpha}A^{b}_{\mu})(\partial_{\beta}c^{c})
\right) \nonumber \\
&& + O(\theta^{2})\nonumber
\end{eqnarray}
where $S$ is the gauge-fixed commutative invariant action, 
and $s^{(\theta)}$ is the noncommutative BRS operator, which, 
at first order in
$\theta$, is
\begin{eqnarray}
    s^{(\theta)} \phi^{a} &=& s \phi^{a} + 
    \half d^{abc}\theta^{\alpha\beta}
    (\partial_{\alpha}\phi^{b})(\partial_{\beta}c^{c})
    \nonumber \\
    s^{(\theta)} A^{a}_{\mu} &=& s A^{a}_{\mu} + 
    \half d^{abc}\theta^{\alpha\beta}
    (\partial_{\alpha}A^{b}_{\mu})(\partial_{\beta}c^{c})
    \nonumber \\
    s^{(\theta)} c^{a} &=& s c^{a} - 
    \frac{1}{4} d^{abc}\theta^{\alpha\beta}
    (\partial_{\alpha}c^{b})(\partial_{\beta}c^{c})
    \label{3.9} \\
    s^{(\theta)} \bar{c}^{a} &=& s \bar{c}^{a}
    \nonumber \\
    s^{(\theta)} b^{a} &=& s b^{a} 
    \nonumber
    \end{eqnarray}
 It is highly nontrivial that the two basic features 
 of the BRS operator \equ{2.8} and \equ{2.9} are conserved when 
 the ordinary product is deformed, via the introduction of the
 $\theta$ parameter, into the Moyal one:
 \begin{eqnarray}
     s^{(\theta)} S^{(\theta)} &=& 0\label{3.10} \\
     (s^{(\theta)})^{2} &=& 0\label{3.11}
     \end{eqnarray}
     It is far less obvious that the deformation of the ordinary
     product into the Moyal one is the only way which allows the
     introduction of the $\theta$ parameter in a way which leads to an
     action symmetric under a nilpotent operator.
     
     It can be verified that the action $S^{(\theta)}$ \equ{3.8} keeps the
     symmetries of its commutative counterpart $S$, namely it 
     is invariant under the supersymmetry $\delta_{\mu}$ \equ{2.16}
     \eq
     \delta_{\mu} S^{(\theta)} = 0\ ,
     \eqn{3.12}
     and satisfies the ghost equation \equ{2.21} as well
     \eq
     {\cal G}^{a} S^{(\theta)} = 0\ .
     \eqn{3.13}
     It is not surprising that the above two symmetries remains
     unaltered although the action is non trivially modified by the
     introduction of $\theta^{\mu\nu}$. As already pointed out, the
     symmetry $\delta_{\mu}$ is related to the topological character of
     the theory, which is not dismantled by $\theta^{\mu\nu}$ which,
     in two dimensions, is proportional to the Lorentz invariant
     Levi--Civita tensor $\epsilon^{\mu\nu}$. On the other hand, the
     identity \equ{3.13} is well understood once we notice that in the
     Moyal product \equ{3.1} the ghost field $c^{a}(x)$ appears
     in the $\theta$-dependent part of the action $S^{(\theta)}$ only 
     differentiated, hence unaffected by the operator ${\cal G}^{a}$.
     
     The superalgebra \equ{2.17} survives in the noncommutative case
     \begin{eqnarray}
    ({s^{(\theta)}})^{2}&=&0 \nonumber\\
    \{\delta_{\mu},\delta_{\nu}\} &=& 0\label{3.14} \\
    \{\delta_{\mu},s^{(\theta)}\} &=& \partial_{\mu} + \mbox{eqs of motion}
    \nonumber\end{eqnarray}
More in detail, the last of \equ{3.14} reads
\begin{eqnarray}
    \{\delta_{\mu},s^{(\theta)}\}A^{a}_{\nu} &=& \partial_{\mu}A^{a}_{\nu}
    +\epsilon_{\mu\nu}\fud{S^{(\theta)}}{\phi^{a}}\label{3.15} \\
     \{\delta_{\mu},s^{(\theta)}\}\phi^{a} &=&
     \partial_{\mu}\phi^{a}
    +\epsilon_{\mu\nu}\fud{S^{(\theta)}}{A^{a}_{\nu}}\label{3.16} \\
     \{\delta_{\mu},s^{(\theta)}\}c^{a} &=& \partial_{\mu}c^{a} 
     \label{3.17}\\
      \{\delta_{\mu},s^{(\theta)}\}{\bar c}^{a} &=&
      \partial_{\mu}{\bar c}^{a}\label{3.18} \\
       \{\delta_{\mu},s^{(\theta)}\}b^{a} &=& \partial_{\mu}b^{a} 
       \label{3.19}
       \end{eqnarray}
       Since the BRS operator is modified from $s$ to $s^{(\theta)}$, 
       the source term in the action is accordingly deformed, at the
       first order in $\theta$, into
       \begin{eqnarray}
       S^{(\theta)}_{ext} &=& S_{ext} + 
       \frac{1}{2}\theta^{\alpha\beta}d^{abc}\int d^{2}x\ \left (      
       {A^{*}}^{a\mu}(\partial_{\alpha}A^{b}_{\mu})(\partial_{\beta}c^{c})
       \right. \label{3.20}\\
       &&
       \left.
       +\
{\phi^{*}}^{a}(\partial_{\alpha}\phi^{b})(\partial_{\beta}c^{c})
-\frac{1}{2} 
{c^{*}}^{a}(\partial_{\alpha}c^{b})(\partial_{\beta}c^{c})\right )
\nonumber
\end{eqnarray}

Summarizing, the noncommutative, two dimensional classical $BF$ action
\eq
\S^{(\theta)}=S^{(\theta)}_{inv} +S^{(\theta)}_{gf}+S^{(\theta)}_{ext}
\eqn{3.21}
although non trivially modified by the introduction of the
noncommutative parameter $\theta^{\mu\nu}$, is characterized by the
same set of symmetries displaying the same algebra as its commutative
counterpart, namely the Slavnov-Taylor identity
\eq
{\cal S}(\S^{(\theta)}) =0\ ,
\eqn{3.22}
the supersymmetry Ward identity
\eq
W_{\mu}\S^{(\theta)}=\D_{\mu}\ ,
\eqn{3.23}
the ghost equation
\eq
{\cal G}^{a}\S^{(\theta)} = \D^{a}\ ,
\eqn{3.24}
and the Landau gauge condition
\eq
\fud{\S^{(\theta)}}{b^{a}} = \partial_{\mu}A^{a\mu}\ .
\eqn{3.25}
We stress again the non triviality of the existence of some of the
above symmetries for the $\theta$-modified action. Particularly
remarkable is the existence of a noncommutative, nilpotent, BRS symmetry
$s^{(\theta)}S^{(\theta)} =0$ and of the supersymmetry
$\delta_{\mu}S^{(\theta)}=0$.

\section{Stability}

A necessary condition for the renormalizability of a quantum field
theory is the stability of the classical action under radiative
corrections. In general, stability is achieved if, after perturbing the classical 
action $\S$ with an infinitesimal functional with the same quantum
numbers as $\S$
\eq
\S \longrightarrow \S + \epsilon\S^{(c)}
\eqn{4.1}
and after imposing that the perturbed action satisfies the same set of
constraints on $\S$, the outcome is that the perturbation $\S^{(c)}$ can
be reabsorbed in $\S$ through a redefinition of the fields and
parameters of the theory. In particular, no new functional monomials,
with respect to the classical action $\S$, should survive this process.
Otherwise, the theory would not be stable, and no
renormalizations could be invoked in order to reabsorb radiative
corrections. The theory would lose its predictive power and hence
would not be renormalizable. 

The renormalizability of the commutative,
two-dimensional $BF$ model has been proven in \cite{Blasi:1992hq}. The
supersymmetry $\d_{\mu}$ is crucial for the stability of the theory,
for it prevents that the coefficients of each monomial appearing in 
the action depend on infinite polynomials in the scalar field 
$\phi^{a}(x)$~:
\eq
{\cal F}(\phi) = D^{a_{1}\ldots a_{n}}\ 
\phi^{a_{1}}\ldots\phi^{a_{n}}
\eqn{4.2}
where $D^{a_{1}\ldots a_{n}}$ are completely symmetric invariant
tensors. Monomials like \equ{4.2} are gauge invariant, and do not affect
power counting prescriptions, since the scalar field
$\phi^{a}(x)$ is dimensionless in two spacetime dimensions.
Nonetheless, those infinite set of field dependent coefficients 
are not present in the classical action $\S$, which 
would therefore be unstable and non renormalizable if the
supersymmetry $\d_{\m}$ would not occur. Indeed, the terms \equ{4.2} are
not invariant under the action of the operator $\d_{\m}$: 
\eq
\d_{\m} {\cal F}(\phi)
=
n\ D^{a_{1}\ldots a_{n}}\ 
\e_{\m\n}\ (\partial^{\n}\bar{c}^{a_{1}})\
\phi^{a_{2}}\ldots\phi^{a_{n}}
\eqn{4.3}
and hence are forbidden. On the other hand, polynomials in color
singlet built with the field strength $F^{a}_{\m\n}(x)$, like for
instance
\eq
\int d^{2}x F^{a}_{\m\n}F^{a\m\n}\ ;\ 
\int d^{2}x\ d^{abc}{\widetilde F}^{a}{\widetilde F}^{b}{\widetilde
F}^{c}\ ; \ldots
\eqn{4.4}
where ${\widetilde F}^{a}(x)$ = $\e^{\m\n}F^{a}_{\m\n}(x)$, are
invariant both under BRS and $\d_{\m}$ symmetries, but violate
power counting (notice that, in two dimensions 
${\widetilde F}^{a}{\widetilde F}^{a} = 2\ F^{a}_{\m\n}F^{a\m\n}$).

The introduction of $\theta^{\m\n}$ drastically changes this scenario. The
general form of the perturbation $\S^{(c)}$ in \equ{4.1} now takes the form
\eq
\S^{(c)}\longrightarrow \S^{(c)} + \S^{(c,\theta)}
\eqn{4.5}
where
\eq
\S^{(c,\theta)} \equiv \theta^{\m\n}\S_{\m\n}
+\theta^{\m\n}\e^{\r\s}\S_{\m\n\r\s}
+\theta^{\m\n}\e_{\m\n}\S
+\theta^{\m\a}\e_{\a}^{\n}\S^{\prime}_{\m\n}
\eqn{4.6}
and $\S_{\m\n},\S_{\m\n\r\s},\S$ and $\S^{\prime}_{\m\n}$ do
not depend on the Levi-Civita tensor $\e^{\a\b}$, and have canonical 
dimensions 4 and Faddeev-Popov
charge 0. Now, due to the fact that, in two dimensions,
$\theta^{\m\n}$ is proportional to $\e^{\m\n}$ \equ{3.4}, the counterterm
finally reads
\eq
\S^{(c,\theta)}= \frac{1}{\theta^{2}}
\left (
\e^{\m\n}X_{\m\n} + X
\right )
\eqn{4.7}
with $X_{\m\n}$ and $X$ not depending on $\e^{\m\n}$.

The perturbed action must satisfy the same constraints as
$\S^{(\theta)}$. This implies, at the first order in the perturbation 
parameter $\e$
\begin{eqnarray}
    \fud{\S^{(c,\theta)}}{b^{a}} &=& 0 \label{4.8}\\
    {\cal G}^{a}\S^{(c,\theta)} &=& 0 \label{4.9}\\
    W_{\r}\S^{(c,\theta)} &=& 0 \label{4.10}\\
    B_{\S^{(\theta)}}(\S^{(c,\theta)}) &=& 0\label{4.11}\ `
    \end{eqnarray}
    where $B_{\S^{(\theta)}}$is the linearized Slavnov-Taylor operator
\begin{eqnarray}
B_{\S^{(\theta)}}=
\Tr \int d^{2}x && \!\!\!\!\left ( 
\fud{\S^{(\theta)}}{A^{*\mu}} \fud{}{A_{\mu}} +
\fud{\S^{(\theta)}}{A_{\mu}}\fud{}{A^{*\mu}}  +
\fud{\S^{(\theta)}}{\phi^{*}} \fud{}{\phi} +
\fud{\S^{(\theta)}}{\phi}\fud{}{\phi^{*}}  + \right. \nonumber \\
&&\left. 
\fud{\S^{(\theta)}}{c^{*}} \fud{}{c} +
 \fud{\S^{(\theta)}}{c}\fud{}{c^{*}}
+ b\fud{}{\bar{c}}
\right )\label{4.12}
\end{eqnarray}
The algebra formed by the supersymmetry Ward operators $W_{\m}$ 
\equ{2.19} and 
the Slavnov-Taylor operator $B_{\S^{(\theta)}}$ \equ{4.12} is
\begin{eqnarray}
    (B_{\S^{(\theta)}})^{2} &=& 0 \label{4.13}\\
    \{W_{\m},W_{\n}\}     &=& 0  \label{4.14}\\
    \{B_{\S^{(\theta)}},W_{\m}\} &=& {\cal P}_{\m}
    \equiv \sum_{\mbox{all fields $\Phi$}} \int d^{2}x\ 
    \partial_{\m}\Phi\fud{}{\Phi}\label{4.15}
\end{eqnarray}

The first two conditions \equ{4.8} and \equ{4.9} 
are satisfied by a functional which does not
depend on the Lagrange multiplier $b^{a}(x)$ and depends on the
undifferentiated ghost
field $c^{a}(x)$ at most once\footnote{To take into account, for 
instance, terms like $\int d^{2}x\ 
\epsilon^{\m\n}d^{abc}c^{a}\partial_{\m}c^{b}\partial_{\n}c^{c}$, 
which satisfies, indeed, the ghost equation \equ{4.9}.}. Moreover, 
as a consequence of the gauge condition \equ{3.25} and of the
Slavnov-taylor identity \equ{3.22}, the action $\S^{(\theta)}$, like any
other gauge theory,  automatically satisfies an 
additional symmetry, called the antighost equation 
\eq
\left (\fud{}{\bar{c}^{a}} + \partial^{\m}\fud{}{A^{\star a\m}}
\right )\S^{(\theta)} =0
\eqn{4.16}
which is satisfied if the fields $A^{\star a \m}$ and $\bar{c}^{a}$
appear in the action only through the combination 
\eq
\widehat{A}^{\star a\m} \equiv {A}^{\star a\m} +
\partial^{\m}\bar{c}^{a}
\eqn{4.17}
Concerning the supersymmetry condition \equ{4.10}, its most general 
solution is
\eq 
\Sigma^{(c,\theta)}=W_{\m}W_{\n}\S^{\m\n},
\eqn{4.18}
where $\S^{\m\n}=-\S^{\n\m}$ is a kind of ``prepotential'', in close 
analogy to what happens in $N=2$ Super Yang-Mills theory \cite{Blasi:2000qw}. 

In order to show this, we develop Lorentz indices:
Eq. \equ{4.18} reads
\eq
\Sigma^{(c,\theta)}=2\ W_{1}W_{2}\S^{12}
\eqn{4.18bis}
where
\begin{eqnarray}
W_{1} &=& \int d^{2}x\ \left(
\widehat{A}^{\star a}_{2}\fud{}{\phi^{a}} +
\phi^{\star a} \fud{}{A^{a}_{2}} +
A^{a}_{1} \fud{}{c^{a}}  -
c^{\star a} \fud{}{\widehat{A}^{\star a}_{1}}
\right) \label{4.19}\\
W_{2} &=& \int d^{2}x\ \left(
-\widehat{A}^{\star a}_{1}\fud{}{\phi^{a}} -
\phi^{\star a} \fud{}{A^{a}_{1}} +
A^{a}_{2} \fud{}{c^{a}}  -
c^{\star a} \fud{}{\widehat{A}^{\star a}_{2}}
\right)\label{4.20}
\end{eqnarray}
Therefore, the corresponding adjoint operators read
\begin{eqnarray}
W^{\dagger}_{1} &=& \int d^{2}x\ \left(
\phi^{a}\fud{}{\widehat{A}^{\star a}_{2}} +
A^{a}_{2} \fud{}{\phi^{\star a}} +
c^{a} \fud{}{A^{a}_{1}}  -
\widehat{A}^{\star a}_{1} \fud{}{c^{\star a}}
\right) \label{4.21}\\
W^{\dagger}_{2} &=& \int d^{2}x\ \left(
-\phi^{a}\fud{}{\widehat{A}^{\star a}_{1}} -
A^{a}_{1} \fud{}{\phi^{\star a}} +
c^{a} \fud{}{A^{a}_{2}}  -
\widehat{A}^{\star a}_{2}\fud{}{c^{\star a} }
\right)\label{4.22}
\end{eqnarray}
The algebra \equ{4.14} reads
\eq
W_{1}^{2} = W_{2}^{2} = 0\ \ ;\ \ 
\{W_{1},W_{2} \} =0
\eqn{4.23}
and
\eq
(W^{\dagger}_{1})^{2}=(W^{\dagger}_{2})^{2} = 0\ \ ;\ \
\{W_{1},W^{\dagger}_{2}\}=\{W_{2},W^{\dagger}_{1}\} =0
\eqn{4.24}
\eq
\{W^{\dagger}_{1},W_{1}\}=\{W^{\dagger}_{2},W_{2}\}=
\sum_{\mbox{all fields $\Phi$}}\int d^{2}x\ \Phi\fud{}{\Phi}
\equiv {\cal N}
\eqn{4.25}
The operator ${\cal N}$ counts  the number $n_{\Phi}$ of fields $\Phi$
appearing in a generic functional ${\cal F}(\Phi)$
\eq
{\cal N}{\cal F}(\Phi) =
\sum_{\mbox{all fields $\Phi$}} n_{\Phi}\ {\cal F}(\Phi)
\equiv N\ {\cal F}(\Phi)
\eqn{4.26}
We stress that $W_{1}$ and $W_{2}$, like their adjoints, 
are nilpotent operators with vanishing 
local cohomology, since all fields appear as BRS doublets \cite{Piguet:1995er}. 

Now, we are interested in the most general local integrated 
functional $X$ satisfying
\eq
W_{1}X^{(p)}_{(q)}=W_{2}X^{(p)}_{(q)}=0
\eqn{4.27}
where $p$ and $q$ are mass dimension and ghost number of the 
functional $X^{(p)}_{(q)}$, respectively.
Our aim is to show that any solution of Eq \equ{4.27} can 
be written as follows
\eq
X^{(p)}_{(q)} = W_{1}W_{2} Y^{(p-2)}_{(q+2)}
\eqn{4.28}
Notice indeed that, from the expression \equ{2.19}, the operators 
$W_{\m}$ raise by one unit the mass dimension and lower by one unit 
the Faddeev-Popov charge.

Let us prove the result \equ{4.28}.

From \equ{4.27}
we have that 
\eq
X^{(p)}_{(q)}=W_{1}X^{(p-1)}_{(q+1)}=W_{2}\widehat{X}^{(p-1)}_{(q+1)}
\eqn{4.30}
where $X^{(p-1)}_{(q+1)}$ and $\widehat{X}^{(p-1)}_{(q+1)}$ 
are generic functionals.
Hence it holds
\eq
W^{\dagger}_{1}W_{1}X^{(p-1)}_{(q+1)}=
W^{\dagger}_{1}W_{2}\widehat{X}^{(p-1)}_{(q+1)}
\eqn{4.31}
Therefore, using \equ{4.25} and \equ{4.26},
\eq
{\cal N}X^{(p-1)}_{(q+1)}-W_{1}W^{\dagger}_{1}X^{(p-1)}_{(q+1)}=
-W_{2}W^{\dagger}_{1}\widehat{X}^{(p-1)}_{(q+1)}
\eqn{4.32}
From \equ{4.26}, we can write
\eq
X^{(p-1)}_{(q+1)}=\frac{1}{N}\left (W_{1}W^{\dagger}_{1}X^{(p-1)}_{(q+1)}
-W_{2}W^{\dagger}_{1}\widehat{X}^{(p-1)}_{(q+1)}\right)
\eqn{4.33}
We then conclude that
\eq
X^{(p)}_{(q)}=W_{1}X^{(p-1)}_{(q+1)}=W_{1}W_{2}Y^{(p-2)}_{(q+2)}
\eqn{4.34}
(where $Y^{(p-2)}_{(q+2)} = 
-\frac{1}{N}W^{\dagger}_{2}\widehat{X}^{(p-1)}_{(q+1)}$),
which is the desired result \equ{4.28}, or \equ{4.18bis}.

Let us summarize our knowledge on the prepotential $\S^{\m\n}$ 
appearing in \equ{4.18}:
\begin{itemize}
\item It is a local integrated functional, with canonical dimension 
      and Faddeev-Popov charge $+2$;
\item It may depend on the quantum fields $A^{a}_{\m}(x)$, 
      $\phi^{a}(x)$, $\partial_{\m}c^{a}(x)$, and at most once on the 
      undifferentiated ghost field $c^{a}(x)$;
\item It may depend on the external sources $\A(x)$, $c^{\star a}(x)$.
      and $\phi^{\star a}(x)$
\end{itemize}
Making explicit the dependence on the Levi-Civita tensor $\e_{\m\n}$, 
as we did in \equ{4.7}, the most general candidate for $\S_{\m\n}$ is
\eq
\S_{\m\n} =\frac{1}{\theta^{2}}( \e_{\m\n}X+X_{\m\n})
\eqn{4.35}
with
\begin{eqnarray}
X &=& \int d^{2}x\ c^{a}\left(
      \partial^{2}c^{b}R^{ab}_{1}(\phi) + \partial^{\m}c^{b} 
      R^{ab}_{2\m}(\phi,A) \right )\label{4.36} \\
X_{\m\n} &=& \int d^{2}x\ c^{a}\left(
      \partial_{\m}c^{b} R^{ab}_{\n}(\phi,A) -  
      \partial_{\n}c^{b} R^{ab}_{\m}(\phi,A) \right)\label{4.37}
\end{eqnarray}
where $R^{ab}_{1}(\phi)$, $R^{ab}_{2\m}(\phi,A)$ and $R^{ab}_{\m}(\phi,A)$
are generic functions. 

Since, due to \equ{4.8}, $X_{\m\n}$ and $X$ are independent from the 
Lagrange multiplier $b^{a}(x)$, recalling \equ{2.21},
the ghost equation condition \equ{4.9} reads
\begin{eqnarray}
    {\cal G}^{a}X &=& \int d^{2}x\ \fud{}{c^{a}} X = 0\label{4.38} \\
    {\cal G}^{a}X_{\m\n} &=& \int d^{2}x\ \fud{}{c^{a}} X_{\m\n} = 0
    \label{4.39}
\end{eqnarray}
which give, respectively
\begin{eqnarray}
    \int d^{2}x\ \partial^{\m}c^{b}(-\partial_{\m}R^{ab}_{1}(\phi) + 
    R^{ab}_{2\m}(\phi,A)) &=& 0\label{4.40} \\
    \int d^{2}x\ c^{b}(-\partial_{\m}R^{ab}_{\n}(\phi,A) + 
    \partial_{\n}R^{ab}_{\m}(\phi,A)) &=& 0\label{4.41}
\end{eqnarray}
which are satisfied if
\begin{eqnarray}
    R^{ab}_{2\m}(\phi,A)) &=& \partial_{\m}R^{ab}_{1}(\phi)\label{4.42} \\
    R^{ab}_{\m}(\phi,A) &=& \partial_{\m}R^{ab}(\phi)\label{4.43}
\end{eqnarray}
where again $R^{ab}(\phi)$ is a generic function. Notice that no dependence 
on the gauge field $A^{a}_{\m}(x)$ is allowed in the prepotential 
$\S_{\m\n}$, reminding again $N=2$ Super Yang-Mills theory \cite{Blasi:2000qw}.
Hence, we have
\begin{eqnarray}
    X &=& \int d^{2}x\ \partial^{\m}c^{a}\partial_{\m}c^{b}\ 
    R^{ab}_{1}(\phi)\label{4.44} \\
    X_{\m\n} &=& \int d^{2}x\ \partial_{\m}c^{a}\partial_{\n}c^{b} \
    R^{ab}(\phi)\label{4.45}
\end{eqnarray}
where $R^{ab}_{1}(\phi)$ and $R^{ab}(\phi)$ (which we rename $R^{ab}_{2}(\phi)$) 
are generic functions of the scalar field $\phi^{a}(x)$, and 
$R^{ab}_{1}(\phi) = - R^{ba}_{1}(\phi)$, 
$R^{ab}_{2}(\phi) = R^{ba}_{2}(\phi)$.

Therefore, our candidate for the $O(\theta)$ counterterm is
\eq
\S^{(c,\theta)} = W_{\m}W_{\n} \int d^{2}x\ \frac{1}{\theta^{2}}\left (
\e^{\m\n} \partial^{\r}c^{a} \partial_{\r}c^{b}\ R^{ab}_{1}(\phi) +
\partial^{\m}c^{a}\partial^{\n}c^{b}\ R^{ab}_{2}(\phi) \right)
\eqn{4.46}
On integrated local functionals $\int d^{2}x\ 
F(\Phi)$, the algebraic relation \equ{4.15} reads
\eq
\{B_{\S^{(\theta)}},W_{\m}\} \int d^{2}x\ F(\Phi) = 
{\cal P}_{\m} \int d^{2}x\ F(\Phi) =
\int d^{2}x\ \partial_{\m}F(\Phi) = 0
\eqn{4.48}
and the Slavnov-Taylor constraint \equ{4.11} becomes
\footnote{The prepotential $\S_{\m\n}$ does not depend on the external 
sources, and the action of the linearized Slavnov-Taylor operator 
\equ{4.12} on the quantum fields coincides with that of the BRS operator 
$s$, since 
$B_{\S^{(\theta)}}\Phi=\fud{\S^{(\theta)}}{\Phi^{\star}}=s\Phi$, 
where $\Phi$ is a generic 
quantum field $\Phi=(A,c,\phi)$.}
\eq
B_{\S^{(\theta)}}\S_{\m\n} = s \S_{\m\n} =0
\eqn{4.47}

Therefore, it must be
\eq
s\ \int d^{2}x\ \left (
\e^{\m\n} \partial^{\r}c^{a} \partial_{\r}c^{b}\ R^{ab}_{1}(\phi) +
\partial^{\m}c^{a}\partial^{\n}c^{b}\ R^{ab}_{2}(\phi) \right) =0
\eqn{4.49}
and it is easily seen that this constraint is satisfied if
\begin{eqnarray}
sR^{ab}_{1}(\phi) &=& f^{amn}c^{m}R^{bn}_{1}(\phi) - 
                      f^{bmn}c^{m}R^{an}_{1}(\phi)\label{4.50} \\
sR^{ab}_{2}(\phi) &=& - f^{amn}c^{m}R^{bn}_{2}(\phi) - 
                        f^{bmn}c^{m}R^{an}_{2}(\phi)\label{4.51}
\end{eqnarray}    
Locality imposes that $R^{ab}_{1}(\phi)$ and $R^{ab}_{2}(\phi)$ are 
not really generic functions, but polynomials in $\phi^{a}(x)$. Taking 
into account the symmetries in the color indices $a$ and $b$, we can write
\begin{eqnarray}
R^{ab}_{1}(\phi) &=& \a f^{abc}\phi^{c} + O(\phi^{2})
=\sum^{\infty}_{n=1} T^{[ab]a_{1}..a_{n}}_{1}\phi^{a_{1}}..\phi^{a_{n}}
\nonumber \\
&=&\sum^{\infty}_{n=1}T^{[ab]P_{n}}_{1}\Phi^{P_{n}}\label{4.52} \\
R^{ab}_{2}(\phi) &=& \b\d^{ab} + \g d^{abc}\phi^{c} + O(\phi^{2})
=\sum^{\infty}_{n=0} T^{(ab)a_{1}..a_{n}}_{2}\phi^{a_{1}}..\phi^{a_{n}}
\nonumber \\
&=& \sum^{\infty}_{n=0}T^{(ab)P_{n}}_{2}\Phi^{P_{n}}\label{4.53}
\end{eqnarray}
where $\a$,$\b$ and $\g$ are constants, and we adopted the short-hand notation 
\mbox{$P_{n}\equiv 
(p_{1}\ldots p_{n})$}. More in general, rigid gauge 
invariance requires that $T^{abP_{n}}_{1,2}$ are constant invariant 
tensors \cite{Piguet:1995er}, built from the Kronecker $\delta^{ab}$, the gauge 
group structure constants $f^{abc}$ and the completely symmetric rank 
three tensor $d^{abc}$. Therefore, conditions \equ{4.50} and \equ{4.51} are 
automatically satisfied, and the most general counterterm, satisfying 
all the constraints \equ{4.8} - \equ{4.11}, is
\begin{eqnarray}
\S^{(c,\theta)} &=& W_{\m}W_{\n} \int d^{2}x\ \frac{1}{\theta^{2}}\left (
\sum^{\infty}_{n=1}T^{[ab]P_{n}}_{1}\ \e^{\m\n}c^{a\r}c^{b}_{\r}\Phi^{P_{n}}
+
\sum^{\infty}_{n=0}T^{(ab)P_{n}}_{2}\
c^{a\m}c^{b\n}\Phi^{P_{n}}\right)
\nonumber \\
&\equiv&
\S^{(c,\theta)}_{1} + \S^{(c,\theta)}_{2}
\label{4.54}\end{eqnarray}
with
\begin{eqnarray}
\S^{(c,\theta)}_{1} &=& \sum^{\infty}_{n=1}\int d^{2}x\ 
T^{[ab](a_{1}..a_{n})}_{1} 
\left [
4\partial^{\m}\phi^{\star a}c^{b}_{\m}\phi^{a_{1}}\phi^{a_{2}} +
2\e^{\m\n}\partial^{\r}A^{a}_{\n}
\partial_{\r}A^{b}_{\m}\phi^{a_{1}}\phi^{a_{2}} 
\right.
\nonumber \\
&&
\left .
-4n\partial^{\m}A^{a\n}c^{b}_{\m}\widehat{A}^{\star a_{1}}_{\n}\phi^{a_{2}} +
4n c^{\a\m}c^{b}_{\m}c^{\star a_{1}}\phi^{a_{2}} 
\right.
 \label{4.55}\\
&& \left.
+\e^{\m\n}n(n-1)c^{a\rho}c^{b}_{\rho}\widehat{A}^{\star a_{1}}_{\m}\widehat{A}^{\star a_{2}}_{\n}
\right ]
\phi^{a_{3}}..\phi^{a_{n}} \nonumber
\\
\S^{(c,\theta)}_{2} &=& \sum^{\infty}_{n=0}
 \int d^{2}x\ T^{(ab)(a_{1}..a_{n})}_{2} 
\left [
2\e^{\m\n}\partial_{\m}\phi^{\star a}c^{b}_{\n}\phi^{a_{1}}\phi^{a_{2}} +
\partial^{\m}A^{a\n}\partial_{\n}A^{b}_{\m}\phi^{a_{1}}\phi^{a_{2}} 
\right.
\nonumber \\
&& \left.
-\partial A^{a}\partial A^{b}\phi^{a_{1}}\phi^{a_{2}} -
2n\e^{\m\a}\partial_{\m}A^{a}_{\n}c^{b\n}\widehat{A}^{\star a_{1}}_{\a}\phi^{a_{2}} +
2n \e^{\m\a}c^{a}_{\m}\partial A^{b} \widehat{A}^{\star a_{1}}_{\a}\phi^{a_{2}} 
\right.
\nonumber \\
&& \left.
+n\e^{\m\n}c^{a}_{\m}c^{b}_{\n}c^{\star a_{1}}\phi^{a_{2}} +
n(n-1)c^{a\m}c^{b\n}\widehat{A}^{\star a_{1}}_{\m}\widehat{A}^{\star 
a_{2}}_{\n}
\right ]
\phi^{a_{3}}..\phi^{a_{n}}\label{4.56}
\end{eqnarray}
where $c^{a}_{\m}\equiv\partial_{\m}c^{a}$. We stress that the 
counterterm \equ{4.54} does not belong to the integrated cohomology of 
the Slavnov-Taylor operator $B_{\S^{(\theta)}}$, since 
it can be written as an exact BRS cocycle:
\eq
\S^{(c,\theta)} = B_{\S^{(\theta)}} Z
\eqn{4.57}
with
\eq
Z=W_{\m}W_{\n}\int d^{2}x\ \frac{1}{\theta^{2}}\left(
\sum_{n=1}^{\infty}\e^{\m\n}T^{[ab]P_{n}}_{1}A^{a\r}c^{b}_{\r}
+
\sum_{n=0}^{\infty}T^{(ab)P_{n}}_{2}A^{a\m}c^{b\n}
\right )\Phi^{P_{n}}
\eqn{4.58}
The counterterm $\S^{(c,\theta)}$ represents the quantum 
corrections of the classical action $\S^{(\theta)}$ \equ{3.21}, 
whose $\theta$-dependent part is
\begin{eqnarray}
\left. \S^{(\theta)} \right |_{O(\theta)} &=&
\frac{1}{2} d^{abc} \theta^{\a\b} \int d^{2}x\ \left (
-\frac{1}{2}\epsilon^{\mu\nu}
\phi^{a}(\partial_{\alpha}A^{b}_{\mu})(\partial_{\beta}A^{c}_{\nu})
+
(\partial_{\mu}\bar{c}^{a})(\partial_{\alpha}A^{b}_{\mu})
(\partial_{\beta}c^{c}) +
\right.
\nonumber \\
&& \left. 
 {A^{*}}^{a\mu}(\partial_{\alpha}A^{b}_{\mu})(\partial_{\beta}c^{c})
       +\
{\phi^{*}}^{a}(\partial_{\alpha}\phi^{b})(\partial_{\beta}c^{c})
-\frac{1}{2} 
{c^{*}}^{a}(\partial_{\alpha}c^{b})(\partial_{\beta}c^{c})
\right )
\label{4.59}\end{eqnarray}
The theory is stable under radiative corrections if the most general 
counterterm can be reabsorbed by a redefinition of fields and 
parameters of the classical theory. In the commutative case, the two 
dimensional BF model is finite, namely no local counterterm is compatible 
with the constraints \equ{4.8} - \equ{4.11} and there is nothing to be 
reabsorbed, or, in other words, no renormalizations of fields or 
parameters are present. On the contrary, it is apparent from the 
direct comparison of the two terms \equ{4.55} and \equ{4.56} forming the 
counterterm and the $\theta$-dependent part of the 
classical action \equ{4.59}, that such reabsorption is, in the noncommutative 
case, impossible. The theory is highly 
unstable, and
radiative corrections are out of control. 

In order that the theory 
is renormalizable, it is not difficult to see that the counterterm, 
instead of consisting in
a double infinity of terms, should collapse into one term only 
\eq
T^{[ab]a_{1}..a_{n}}_{1} = 0\ \ \  \forall n
\eqn{4.60}
and
\eq
T^{(ab)a_{1}..a_{n}}_{2} =
\left \{
\begin{array}{ll}
     0 & \forall n \neq 1  \\
 \frac{1}{4}d^{aba_{1}} & \textrm{if}\ n=1 
\end{array}
\right.
\eqn{4.61}

This counterterm might be reabsorbed by a renormalization of the 
$\theta$ parameter. But this is not the case, and the theory, due to 
the infinite terms present in the radiative corrections, has no 
predictive power. Even before asking the question of the presence of 
noncommutative anomalies (the commutative case is of course 
anomaly-free \cite{Blasi:1992hq}), we must conclude that the noncommutative 
theory, taken just as a quantum field theory, is highly non 
renormalizable. We believe that this illness is not peculiar of the 
two dimensional $BF$ model, but it is due to the 
simple recipe of constructing noncommutative theories from 
their commutative counterpart just substituting the ordinary product 
between fields with the groenewold-Moyal one. 

\section{Anomaly}

As it is well known, anomalies are quantum breakings of symmetries. 
Let $\G$ be the functional generator of 1PI Green functions, or, 
equivalently, quantum vertex, quantum or effective action. It holds
\eq
\G = \S^{(\theta)} + O(\hbar)
\eqn{5.1}
where $\S^{(\theta)}$ is the classical action \equ{3.21}. The constraints 
\equ{3.23}, \equ{3.24}, \equ{3.25} and \equ{4.16}, 
being linear, are easily shown to hold also at the 
quantum level: 
\begin{enumerate}
    \item gauge condition 
    \eq 
    \fud{\G}{b^{a}(x)} = \partial A^{a}
    \eqn{5.2}
    \item antighost equation
    \eq
    \left ( \fud{}{A^{\star a} (x)} + \fud{}{\partial_{\m}\bar{c}^{a} (x)} 
    \right)\G =0
    \eqn{5.3}
    \item ghost equation
    \eq
    {\cal G}^{a}\G=\D^{a}
    \eqn{5.4}
    \item supersymmetry
    \eq
    W_{\m}\G=\D_{\m}
    \eqn{5.5}
\end{enumerate}
The only symmetry whose quantum extension should be handled with care 
is the (nonlinear) Slavnov-Taylor identity \equ{3.22}, which {\it a priori} 
is broken
\eq
{\cal S}(\G) = {\cal A}
\eqn{5.6}
by a quantum breaking $\cal A$ which, according to the Quantum Action 
Principle 
\cite{Lowenstein:1971vf,Lowenstein:1971jk,Lam:1972mb,Lam:1973qa,Clark:1976ym}, 
at the lowest nonvanishing order in $\hbar$, is a 
local integrated functional with canonical dimensions $2$ and 
Faddeev-Popov charge $+1$
\eq
{\cal A} = \D^{(2)}_{(1)} + O(\hbar \D^{(2)}_{(1)}) = \int d^{2}x\ 
\D^{(2)}_{(1)}(x) + O(\hbar \D^{(2)}_{(1)})
\eqn{5.7}
The web of algebraic relations \cite{Blasi:1992hq} results in the following 
consistency conditions on $ \D^{(2)}_{(1)} $: 
    \eq 
    \fud{\D^{(2)}_{(1)}}{b^{a}(x)} = 0\ ;
    \eqn{5.8}
    \eq
    \left ( \fud{}{A^{\star a}_{\m} (x)} + \fud{}{\partial_{\m}\bar{c}^{a} (x)} 
    \right)\D^{(2)}_{(1)} =0\ ;
    \eqn{5.9}
    \eq
    {\cal G}^{a}\D^{(2)}_{(1)}=0\ ;
    \eqn{5.10}
    \eq
    W_{\m}\D^{(2)}_{(1)}=0\ ;
    \eqn{5.11}
    \eq
    B_{\S^{(\theta)}}\D^{(2)}_{(1)}=0\ .
    \eqn{5.12}
The last constraint \equ{5.12} is the cohomology problem commonly known as 
Wess-Zumino consistency condition \cite{Wess:1971yu}. 
The constraints \equ{5.8}-\equ{5.12} are 
basically the same that we have already imposed on the counterterm 
$\S^{(c,\theta)}$, with the difference that the functional 
$\D^{(2)}_{(1)}$ belongs to the Faddeev-Popov sector with charge $+1$ 
instead of zero. 

We already know from the previous section that the solution of  the 
first four constraints is a functional $\D^{(2)}_{(1)}$ which does not 
depend on the Lagrange multiplier $b^{a}(x)$, depends on the 
antighost field $\bar{c}^{a}(x)$ and on the external source $A^{\star 
a}_{\m}(x)$ only through the combination $\A(x)$ \equ{4.17}, it may 
depend on the undifferentiated ghost field $c^{a}(x)$ at most once and 
it can be written as
\eq
\D^{(2)}_{(1)} = W_{\m}W_{\n}\D^{(0)\m\n}_{(3)}\ ,
\eqn{5.13}
where $\D^{(0)\m\n}_{(3)}$ is a local integrated functional with 
dimensions zero and Faddeev-Popov charge $+3$. The only possible 
$\theta$-independent term is is
\eq
\D^{(0)\m\n}_{(3)}=\int d^{2}x\ \e^{\m\n}
c^{a}c^{b}c^{c}R^{[abc]}(\phi)
\eqn{5.14}
with $R^{[abc]}(\phi)$ polynomial in $\phi^{a}(x)$ and antisymmetric 
in the color indices $abc$. But this only 
commutative candidate for the anomaly is ruled out by the ghost 
equation \equ{5.10}, and, as we already knew \cite{Blasi:1992hq}, 
the commutative 
theory is anomaly free.

Let us see if the situation is different at the order $\theta$. 
Recalling that $\theta^{\m\n}\propto\e^{\m\n}$, we must look for 
anomalies whose structure is
\eq
\D^{(2)}_{(1)} = W_{\m}W_{\n}\frac{1}{\theta^{2}}
\left (\e^{\m\n}\D^{(2)}_{(3)}
+ \D^{(2)\m\n}_{(1)} \right)\ .
\eqn{5.15}
The only possibilities for functionals depending on the 
undifferentiated ghost field $c^{a}(x)$ at most once, are
\begin{eqnarray}
\D^{(2)}_{(3)} &=& \int d^{2}x\                    
c^{a}\partial^{\r}c^{b}\partial_{\r}c^{c}\ 
R^{a[bc]}_{1}(\phi)\label{5.16} \\
\D^{(2)\m\n}_{(3)} &=& \int d^{2}x\                    
c^{a}\partial^{\m}c^{b}\partial^{\n}c^{c}\ R^{a(bc)}_{2}(\phi) \ ,
\label{5.17}\end{eqnarray}
with $R^{a[bc]}_{1}(\phi)$ and $R^{a(bc)}_{2}(\phi)$ 
polynomials in $\phi^{a}(x)$.

The ghost equation \equ{5.10} is satisfied if
\begin{eqnarray}
    R^{a[bc]}_{1}(\phi) &=& 0\label{5.18} \\
    R^{a(bc)}_{2}(\phi) &=& \a d^{abc}\ ,\label{5.19}
\end{eqnarray}
where $\a$ is a constant. Hence the only candidate for the 
$O(\theta)$ anomaly is
\eq
\D^{(2)}_{(1)} = W_{\m}W_{\n} \int d^{2}x\ 
\frac{1}{\theta^{2}}\a d^{abc}c^{a}\partial^{\m}c^{b}\partial^{\n}c^{c}\ ,
\eqn{5.20}
but $\D^{(2)}_{(1)}$ is not BRS invariant, hence does not satisfy the 
Wess-Zumino consistency condition \equ{5.12}. We conclude that, at least 
at the first order in $\theta$, non commutativity does not introduce 
anomalies, but nevertheless the noncommutative theory is still not 
renormalizable.

\section{Conclusions}

In this paper we considered the noncommutative deformation of the two 
dimensional BF model, which, taken as an ordinary commutative theory, 
is topological, and it shares with all other commutative, 
topological field theories the property of finiteness. This means 
that the 
topological, commutative BF model has a 
particularly simple quantum extension, since, as it has been shown in 
\cite{Blasi:1992hq}, no local counterterm turns out to be 
compatible with the symmetries 
characterizing the classical theory. In the noncommutative extension 
of quantum field theories, a new parameter comes into play, 
$\theta_{\m\n}$, which has mass dimension minus two. The easy recipe to 
treat a field theory living in the noncommutative plane is that of 
describing it through an ordinary, commutative field theory, where the 
product between quantum fields is substituted by the groenewold-Moyal 
star product \equ{3.2}. This is the so called ``Moyal prescription'' 
\cite{Szabo:2001kg,Douglas:2001ba}. 
Now, as it is widely known, the product between quantum fields is one 
of the delicacies of quantum field theory, since distributions at 
coinciding points are not well defined. The Moyal prescription 
just modifies the product between quantum fields, and we believe that 
serious attention should be paid to such an operation. We showed, in 
the simple example of this finite two dimensional theory, that the 
Moyal prescription leads to a meaningless quantum field theory, and we 
believe that the situation could be even worse in more complicated 
quantum field theories, living in higher dimensional spacetime. We 
do not conclude that in general the noncommutative theories do not 
make sense. Our milder claim is that one should be careful in 
applying to them the usual rules of
{\it quantum field theories}, in the sense that we are going to 
explain. 
The main feature of a quantum field theory is {\it locality}. Locality 
requires that the action of a quantum field theory is an integrated 
local functional, moreover, such quantum functional (namely the 
classical action and its 
quantum radiative corrections) must be {\it analytical} in all 
parameters (masses, coupling constants, \ldots) appearing in the theory. 
We limited ourselves just to the analytical sector in $\theta$, aware 
of the fact that the analytical sector does not include the whole noncommutative 
theory. Non commutativity indeed may open a non analytical 
sector in which the ordinary rules of quantum fields theory do not apply. 
Nevertheless, the analytical sector, in which a $\theta$-expansion is 
allowed, does exist, and it is precisely this which we considered, 
limiting even more ourselves to the first order in $\theta_{\m\n}$. 
On non-analiticity introduced by the noncommutative deformation of 
quantum field theories, see for instance 
\cite{Douglas:2001ba,Wulkenhaar:2001sq}.

Summarizing, our playground is quite restricted: it is the noncommutative 
theory taken in its quantum field theoretical sector, 
which must be local and analytical in all its parameters, including 
$\theta_{\m\n}$, and we looked to the first order in its 
$\theta$-expansion. But, even within this fence, we found that the 
quantum corrections are represented by a double infinity of terms, 
\equ{4.54}, which cannot be reabsorbed, as they should for a 
renormalizable quantum field theory, by a redefinition of fields and 
parameters of the theory. Therefore, the model is not renormalizable 
in the sense 
that it looses any predictive power. This occurrence should be 
interpreted, in our opinion, as a signal that there must be some 
novel feature in the noncommutative model in order to make it 
sensible. For instance, the two families of couplings could be a 
Taylor expansion of some functions; but this needs a precise relation 
between the coupling constants, relation which is not enforced by the 
known symmetries. Is there some new symmetry brought about by 
noncommutativity?

At this point, it is useless to go to higher orders in the 
$\theta$-expansion, since we can already conclude that the noncommutative 
theory does not exist as a quantum field theory. 
It is just something else. We believe that the situation 
might be more dramatic in higher dimensions, where the Landau gauge 
is not a compulsory choice, as it is in two and three dimensions, 
where the gauge parameter is massive, and gauge choices else than the 
Landau one lead to severe infrared problems. The Landau gauge theory 
is characterized by the ghost equation condition \equ{5.4}, which 
represents a very strong constraint on radiative corrections 
\cite{Blasi:1990xz}. 
The 
double infinities of instabilities \equ{4.54} would be much more 
without the protection of this constraint. On the other hand, at 
first order in $\theta_{\m\n}$, we did not find any noncommutative 
anomaly. This reinforces the impression that the instability 
problem could be overcome by adding some constraints which are 
peculiar of the noncommutative model.

We are very curious to investigate whether a 
$\theta$-dependent anomaly might occur at higher orders and/or in 
different theories, like for instance four dimensional Yang-Mills 
theory, where we expect that, besides the obvious noncommutative 
extension of the Adler-Bardeen-Jackiw anomaly, other, purely noncommutative, 
anomalies exist.
%
%
%
%
%
%

\end{document}